
\input harvmac
\Title{\vbox{\baselineskip12pt\hbox{CERN-TH-95-201}\hbox{FTUAM-95-101}
\hbox{hep-th/9507112}}}
{\vbox{\centerline{T-Duality and Space-Time Supersymmetry}}}

\centerline{Enrique \'Alvarez\footnote{$^\bullet$}{and: Departamento
de F\'{\i}sica Te\'orica, Universidad Aut\'onoma,
28049 Madrid, Spain}, Luis \'Alvarez-Gaum\'e and Ioannis
Bakas\footnote{$^\dagger$}
{permanent address: Department of Physics, University of Patras,
26110 Patras,
Greece}}

\bigskip\centerline{CERN-TH Division}
\centerline{1211 Geneva 23, Switzerland}

\vskip .3in
We analyze in detail the possible breaking of space-time
supersymmetry
under T-duality transformations. We find that when appropiate
world-sheet
effects are taken into account apparent puzzles concerning
supersymmetry in
space-time are solved. We study T-duality in general heterotic
$\sigma$-models
analyzing possible anomalies, and we find some modifications of
Buscher's
rules. We then work out a simple but representative example which
contains
most of the difficulties. We also consider the issue of supersymmetry
versus
duality for marginal deformations of WZW models and present a
mechanism that
restores supersymmetry dynamically in the effective theory.

\Date{July 1995}


\newsec{Introduction}
Target-space duality (T-duality) \ref\tdual{E. Alvarez, L.
Alvarez-Gaum\'e
and Y. Lozano, Nucl. Phys. (Proc. Supp.) 41 (1995) 1;
 A. Giveon, M. Porrati and
E. Rabinovici,
Phys. Rep. 244 (1994) 77} is an important example of discrete string
symmetries
showing the equivalence of strings propagating on background
space-times with
different geometry and (sometimes) topology. In some recent papers
\ref\tsusy{
I. Bakas, Phys. Lett. B343 (1995) 103; I. Bakas and K. Sfetsos, Phys.
Lett.
B349 (1995) 448;
E. Bergshoeff, R. Kallosh and T. Ort\'{\i}n,
Phys. Rev. D51 (1995) 3009} it was argued that from the point of view
of the
low energy effective action, the original and T-dual theories do not
share the
same number of space-time supersymmetries.
If we identify them with the number of Killing spinors in the given
space-time background \ref\killing{M. Green, J. Schwarz and E.
Witten,
Superstrings, volume 2 (Cambridge University Press, 1987), and
references
therein}, it is not difficult to find examples where such numbers
come
out to be different.
There are several reasons why this conclusion is hard to accept.
First if we take the point of view that a duality transformation is a
change
of variables, or a canonical transformation \ref\canonical{E.
Alvarez,
L. Alvarez-Gaum\'e and Y. Lozano, Phys. Lett. B336 (1994) 183}, the
symmetries
of
the original theory should be preserved although they may become
non-local
\ref\kiritsis{E. Kiritsis, Nucl. Phys. B405 (1993) 109}.
Second, there are general theorems \ref\banks{T. Banks and L. Dixon,
Nucl. Phys. B307 (1988) 93}
relating space-time supersymmetries to symmetries on the world-sheet.
This connection is of a very general nature, and if we can show that
the manipulations involved in carrying out a T-duality transformation
preserve
them, we should expect space-time supersymmetry to be maintained,
although we may have to revise the relationship between world-sheet
properties and those of the the associated low energy effective
action.
\par
T-duality can be defined when the target space geometry admits
isometries.
We can classify the isometries into those with and without fixed
points.
A typical example of isometries without fixed points are the
translational symmetries on tori.
If the isometry has a fixed point at $x_0$, the associated Killing
vector
$k^i$ vanishes at $x_0$. Since $\nabla_i k_j + \nabla_j k_i = 0$, on
the
tangent space at $x_0$, $k^i$ acts via the rotation matrix $\nabla_i
k_j
-\nabla_j k_i $ . In the $\sigma$-model formulation of a string
propagating
on a background, we assume that its size and curvature are
respectively
large and small compared with the inverse string tension $\alpha
^{\prime}$.
Hence, close to the fixed point, $x_0$, we can approximate the target
manifold
by flat space. If for simplicity we consider the case where
coordinates
can be chosen so that $\nabla_i k_j - \nabla_j k_i$ only acts on a
two-plane
$(x,y)$ on the tangent space, locally close to $x_0$ the metric in
the
$(x,y)$ directions takes the form
\eqn\metrica{ds^2 \sim dx^2 + dy^2 \sim d\rho^2 + \rho^2 d\theta^2 ,}
where the Killing vector corresponding to translations in $\theta$ is
$k\sim {\partial
\over \partial \theta}$. Under T-duality \ref\buscher{T. Buscher,
Phys. Lett.
B194 (1987) 51; Phys. Lett. B201 (1988) 466}, \metrica\
is transformed to
\eqn\dualmet{d{\tilde s}^2 \sim d\rho^2 + {\alpha^{\prime}\over
\rho^2}
d {\tilde \theta}^2}
and the dual dilaton receives a contribution $\tilde S = S -{1 \over
2}
\log {\rho^2 \over \alpha^{\prime}}$ . The dual geometry has a
curvature
singularity
at $\rho = 0 $. It is then clear that the usual $\sigma$-model
formulation of strings in this background breaks down close to the
singularity, and more powerful techniques should be used in order to
understand the behavior of strings close to this point.

If one insists
on exploring the metric \dualmet\ from the low energy point of view,
one
easily finds that together with the dilaton above it satisfies the
$\beta$-function
equations to first order in $\alpha^{\prime}$ \ref\betaa{C. Callan,
D. Friedan,
E. Martinec and M. Perry, Nucl. Phys. B262 (1985) 593} .
It is not difficult to verify that
\dualmet\ does not admit Killing spinors, although \metrica\ admits
two.
This is the example studied in the last paper of \tsusy. In this
reference
one considers the heterotic string on flat ten-dimensional Minkowski
space.
This theory has obviously $N=1$ space-time supersymmetry in $D=10$
(with a
Majorana-Weyl spinor generator). If we select two spatial
coordinates,
say ($x, y$), and perform a duality rotation, as in \dualmet, the
supersymmetric
variation of the dilatino does
not vanish \tsusy
\eqn\dilatino{\delta_{\epsilon} \lambda = (\gamma^{\mu}
\partial_{\mu} S)
\epsilon}
and one would be tempted to conclude that the dual theory has no
supersymmetries.
Although this example is particularly simple, it is clear from the
previous
arguments that it is representative of any target space manifold with
a
rotational Killing vector close to its fixed point. Hence we should
understand what happens to space-time supersymmetry on the flat
ten-dimensional
heterotic string under T-duality with respect to one of its
rotational
Killing symmetries. We should also clarify how the apparent puzzle
raised in
\tsusy\  can be understood from the point of view of the low energy
effective
action.
\par
The outline of this paper is as follows. In section 2 we work out the
duality
transformation for a general $(1,0)$ heterotic $\sigma$-model
\ref\ssigma{C.
Hull
and E. Witten, Phys. Lett. 160B (1985) 398; R. Brooks, F. Muhammad
and S.
Gates,
Nucl. Phys. B268 (1986) 599; S. Randjbar-Daemi, A. Salam and J.
Strathdee,
Nucl.
Phys. B320 (1989) 221; G. Moore, P. Nelson , Nucl. Phys. B274 (1986)
509} with
arbitrary
connection and background gauge field. We find that if one does not
want
to have a world-sheet non-local T-dual action due to anomalies which
appear when implementing the duality transformation, one has to
transform
under the isometry the right-moving fermions (which on the heterotic
string
generate the $Spin(32)/ Z_2$ or $SO(16) \times SO(16)$ currents).
This yields a non-trivial transformation of the background gauge
field under
T-duality. We find also that if in the original model the gauge and
the spin
connections match (the simplest and most widely used condition
necessary to
guarantee conformal invariance to order $\alpha^{\prime}$ and anomaly
cancellation on the ten-dimensional effective theory
\ref\anomalias{M. Green,
J. Schwarz, Phys. Lett. B149 (1984) 117; E. Witten, ``Global
Anomalies in
String
Theory'' in: Symposium on Anomalies, Geometry and Topology (edited
by W. Bardeen and A.R. White, World Scientific, 1985); P. Candelas G.
Horowitz,
A.
Strominger and E. Witten,
Nucl. Phys. B258 (1985) 46}), the change in the
gauge field under T-duality ensures the same matching in the dual
theory.
Furthermore if the original theory had $(2,0)$ or $(2,2)$
superconformal
invariance, the dual theory also has these properties; a prerequisite
for the
application of the theorems mentioned above \banks\ concerning
world-sheet
conditions to get space-time supersymmetry.
\par
In section 3 we present how space-time supersymmetry is restored from
the
point of view of the low energy effective action. Here we concentrate
in the
simplest example of the heterotic string considered previously, but
the
generalization to other cases is straightforward.
\par
In section 4 we present a detailed analysis, within the framework of
conformal field theory, of the duality transformations \dualmet, and
show
that there are indeed world-sheet operators on the dual theory
associated
to the space-time supersymmetry charges, although some world-sheet
non-locality is generated. We will find an interesting interplay
between
the picture-changing operator \ref\picture{D. Friedan, E. Martinec
and
S. Shenker, Nucl. Phys. B271 (1986) 93} and T-duality.
\par
In section 5 we consider the issue of supersymmetry versus duality
for
continuous $O(d,d)$ transformations and in particular for marginal
$J \bar{J}$-deformations of a given conformal field theory. We find
that
space-time supersymmetry exhibits the same ``anomalous" behaviour for
generic
values
of the modulus field, when the Killing vector fields have fixed
points in their
action. We also consider a mechanism that restores dynamically the
supersymmetry
in the framework of Kazama-Suzuki models \ref\Kazama{P. Spindel, A.
Sevrin, W.
Troost and
A. van Proeyen, Nucl. Phys. B308 (1988) 662, B311 (1988) 465; Y.
Kazama and H.
Suzuki,
Phys. Lett. B216 (1989) 112; Nucl. Phys. B321 (1989) 232}, and for
$SU(2) \times U(1)$ in particular.
\par
Finally, in section 6 we present our conclusions and directions for
some
further
work.

\newsec{Duality in Heterotic $\sigma$-models}

A heterotic $\sigma$-model is best formulated in (1,0) superspace
\ssigma. (1,0) superfields have the simple form
\eqn\superf{\Phi^i(\sigma,\theta) = x^i (\sigma) + \theta \lambda^i,}
\eqn\superff{\Psi^A(\sigma,\theta) = \psi ^A + \theta F^A.}
We use light-cone coordinates on the world-sheet, $\sigma^{\pm} =
{\sigma^0 \pm
\sigma^1 \over \sqrt 2}$, $\partial_{\pm} = {\partial_0 \pm
\partial_1
\over \sqrt 2}$, and it is useful to introduce the operator:
\eqn\D{ D = {\partial \over \partial \theta} + i \theta {\partial
\over
\partial \sigma^{+}} ; ~~~~
D^2 = i \partial_{+} .}

If we consider a manifold $M$ with metric $g_{ij}$, antisymmetric
tensor
field $b_{ij}$ and a background gauge connection $V_{iAB}$ associated
to
a gauge group $G \in O(32)$ (for simplicity we consider the $O(32)$
heterotic
string), we need two types of superfields: coordinate superfields
$\Phi^i$
\superf\
and gauge superfields. The fermions $\lambda$ and $\psi$ have
opposite
world-sheet chirality, $x^i (\sigma)$ are the fields embedding the
world-sheet
in the target space, and the $F^A$ are auxiliary fields.
The Lagrangian density is given by
\eqn\lagrang{L = \int d\theta \left(-i(g_{ij} + b_{ij}) D\Phi^i
\partial_{-}\Phi^j
- \delta_{AB} \Psi^A {\cal D} \Psi^B \right),}
with
\eqn\covder{{\cal D} \Psi^A = D\Psi^A + V_{i}\,^{A}\,_{B}(\Phi)
D\Phi^i
\Psi^B.}
Eliminating the auxiliary fields one obtains \ssigma
\eqn\lagg{L = (g_{ij} + b_{ij}) \partial_{+} x^i \partial_{-} x^j
+ i g_{ij} \lambda^i D_{-} \lambda^j + i \psi^A  D_{+} \psi^A +
{1 \over 2} F_{ijAB} \lambda^i \lambda^j \psi^A \psi^B,}
with
\eqn\dd{ D_{+} \psi^A = \partial_{+} \psi^A + V_{i}\,^{A}\,_{B}
\partial_{+}
x^i \psi^B ,}
\eqn\ddd{D_{-}\lambda^i = \partial_{-}\lambda^i + (\Gamma^{i}_{jk} +
{1 \over
2}
H^{i}_{jk}) \partial_{-} x^j \lambda^k,}
\eqn\H{H_{ijk} = \partial_i b_{jk} + \partial_j b_{ki} + \partial_k
b_{ij},}
\eqn\F{ F_{ij}\,^{A}\,_{B} = \partial_i V_{j}\,^{A}\,_{B} -
\partial_j
V_{i}\,^{A}\,_{B} + [V_i,V_j]\,^{A}\,_{B}.}
The world-sheet supercurrent is of type (1,0):
\eqn\superc{{\cal G}_{+} = (2 g_{ij} + b_{ij})\partial_{+} x^i
\lambda^j
- {i \over 2} H_{ijk} \lambda^i \lambda^j \lambda^k .}

To carry out a duality transformation in \lagrang\ we need to assume
that the
metric has an isometry under which \lagrang\ and \lagg\ are
invariant.
Following the procedure
outlined in \ref\rv{M. Rocek and E. Verlinde, Nucl. Phys. B373 (1992)
630}
we next gauge the isometry, with
some gauge fields $A_{\pm}$ and add an extra term with a Lagrange
multiplier
making the gauge field strength vanish. If we integrate out the
Lagrange multiplier, $A_{\pm}$ become pure gauge, $A_{\pm} =
\partial_{\pm}
\alpha$.
Using the invariance of the action we can change variables to remove
all
presence of $\alpha$ and recover the original action. If instead we
integrate
first over $A_{\pm}$ and then fix the gauge, we obtain the dual
theory. In
our case we will carry out these steps in a manifestly
(1,0)-invariant
formalism.
The conditions that need to be satisfied by \lagg\ to be able to
gauge an
isometry
have been studied in \ref\hull{C. Hull, Mod. Phys. Lett. A9 (1994)
161}.
If $k^i$ is the Killing
vector of the metric $g_{ij}$, the first term of \lagrang\ is
invariant
provided
\eqn\kh{k^i H_{ijk} = \partial_j v_k - \partial_k v_j}
and
\eqn\deltab{\delta_{k} b_{ij} = \partial_i (k^l b_{lj} + v_j) -
\partial_j (k^l b_{li}).}
The conserved (1,0)-supercurrent for the first term of \lagrang\ is:
\eqn\jotamenos{{\cal J_{-}} = (k_i - v_i) \partial_{-} \Phi^i, ~~~~
{\cal J_{+}} = (k_i + v_i) D\Phi^i,}
so that
$$ D {\cal J_{-}} + \partial_{-} {\cal J_{+}} = 0 .$$

We now introduce (1,0) gauge fields ${\cal A_{-}}$, ${\cal A}$ of
bosonic
and fermionic character respectively
\eqn\amenos{{\cal A_{-}} = A_{-} + \theta \chi_{-}, ~~~~
{\cal A} = \chi + i \theta A_{+}.}
If $\epsilon (\sigma,\theta)$ is the gauge parameter, we can take
\eqn\varmenos{\delta_{\epsilon}{\cal A_{-}} = - \partial_{-} \epsilon
, ~~~~
\delta_{\epsilon}{\cal A} = - D \epsilon .}
Then, to first order in $\epsilon$,
$$(g_{ij} + b_{ij}) D\Phi^i \partial_{-} \Phi^j + {\cal J_{+}}{\cal
A_{-}}
+ {\cal J_{-}}{\cal A}$$
is gauge invariant, with
\eqn\var{\delta_{\epsilon} \Phi^i = \epsilon k^i(\Phi).}
Full gauge invariance can be achieved if we assume that $k^i v_i =
const.$;
in which case the gauge invariant Lagrangian is
$$(g_{ij} + b_{ij}) D\Phi^i \partial_{-} \Phi^j + {\cal J_{+}}{\cal
A_{-}}
+ {\cal J_{-}}{\cal A} + k^2 {\cal A_{-}}{\cal A} .$$

The left-moving part
$$\Psi (D \Psi + V_i D\Phi^i)\Psi$$
is invariant under the global transformation \var\ when the isometry
variation can be compensated by a gauge transformation:
\eqn\dfi{\delta \Phi^i = \epsilon k^i (\Phi),\qquad \delta \Psi = -
\kappa
\Psi ,}
with
\eqn\dev{\delta_k V_i = {\cal D}_i \kappa = \partial_{i} \kappa
+[V_i,\kappa]
,}
which implies
\eqn\kf{k^i F_{ij} = D_j \mu ; ~~~~~
\mu = \kappa - k^i V_i.}
Making $\epsilon$ a function of (1,0) superspace one obtains after
some
algebra:
\eqn\dkin{\delta_{\epsilon} ( \Psi^{T} {\cal D} \Psi) = D\epsilon
\Psi^{T} \mu
\Psi .}

Hence, adding the coupling
$$ {\cal A} \Psi^{T}\mu\Psi , $$
we achieve gauge invariance, because $\Psi^{T}\mu\Psi$ is gauge
invariant:
\eqn\gory{\eqalign{&k^i F_{ij} = \partial_j \mu +[V_j ,\mu],\cr
&0=k^i k^j F_{ij} = k^j \partial_j \mu +[k^j V_j,\mu] = k^j
\partial_j \mu +
[\kappa,\mu].\cr}}
Then,
\eqn\goryy{\delta_{\epsilon}\Psi^{T} \mu \Psi = \Psi^{T} (k^j
\partial_j \mu
+ [\kappa,\mu]) \Psi = 0 .}
The full gauge invariant Lagrangian is:
\eqn\goryyy{L = -i \left( (g_{ij} + b_{ij}) D\Phi^i \partial_{-}
\Phi^j + {\cal
J_{+}}
{\cal A_{-}} + {\cal J_{-}} {\cal A} + k^2 {\cal A_{-}} {\cal A}
\right)
- (\Psi^{T} {\cal D} \Psi + {\cal A} \Psi^{T} \mu \Psi).}

Add now the Lagrange multiplier superfield term
\eqn\lagmult{i \Lambda (D {\cal A_{-}} - \partial {\cal A}).}
Integrating over $\Lambda$ implies that ${\cal A} = D \alpha$,
${\cal A_{-}} = \partial_{-} \alpha$, and using the invariance of
\goryy\ we
obtain
the original theory.
Classical duality is obtained by integrating out ${\cal A}, {\cal
A_{-}}$.
Since \goryy\ is at most quadratic in ${\cal A},{\cal A_{-}}$, we can
solve
their equations of motion to obtain the dual Lagrangian:
\eqn\duall{{\tilde L}_{cl} = -i \left( ({\tilde g}_{ij} + {\tilde
b}_{ij})
D\Phi^i
\partial_{-}\Phi^j + ({\cal J}_{+} + D \Lambda ){1\over k^2}
(\partial_{-}
\Lambda
+ i \Psi^{T} \mu \Psi - {\cal J}_{-}) \right)
- \Psi^{T} {\cal D} \Psi   .}

If we use adapted coordinates to the Killing vector, $k^i {\partial
\over
\partial x^i} = {\partial\over \partial x^0}$, split the coordinates
as $ i =
(0,
\alpha)$ (with $i = 0,1,..., D-1$; $\alpha = 1,...,D-1$),
and choose locally the gauge
$\Phi^0 = 0$, the dual values for ${\tilde g}$, ${\tilde b}$,
${\tilde V}$ are:

\eqn\dualll{\eqalign{&
 {\tilde g}_{00} = {1\over k^2},\cr
&{\tilde g}_{0 \alpha} = {1\over k^2} v_{\alpha},\cr
&{\tilde b}_{0\alpha} = - {1\over k^2} k_{\alpha},\cr
&{\tilde g}_{\alpha\beta} = g_{\alpha\beta} - {k_{\alpha} k_{\beta} -
v_{\alpha}v_{\beta} \over k^2},\cr
&{\tilde b}_{\alpha\beta} = b_{\alpha\beta} +{k_{\alpha} b_{0 \beta}
-
k_{\beta} b_{0 \alpha} \over k^2},\cr
&{\tilde V}_{0 AB} = - {1\over k^2} \mu_{AB},\cr
& {\tilde V}_{\alpha AB} = V_{\alpha AB} - {1\over k^2} (k_{\alpha} +
v_{\alpha}) \mu_{AB}.\cr}}
Since in adapted coordinates we can choose $v_{\alpha} = -
b_{0\alpha}$,
\dualll\
is equivalent to Buscher's formulae \buscher\ ; but we find a change
in
the background gauge field as well. \foot{It came to our attention
that
a similar result was independently derived in \ref\oortin{E.
Bergshoeff,
I. Entrop and R. Kallosh, Phys. Rev D49 (1994) 6663; E. Bergshoeff,
B. Janssen and T. Ortin, ``Solution Generating Transformations and
the
String Effective Action", Groningen preprint UG-1-95, hep-th/9506156,
June 1995} .}
\par
The preceding formulae \dualll\ were obtained using only classical
manipulations.
In general, however, there will be anomalies and the dual action
\duall\ may
not
have the same properties as the original one. Depending on the choice
for $\mu$
and the gauge group $G \in O(32)$, the theory \duall\ may be
afflicted with
anomalies. In this case \duall\ and \goryyy\ are not equivalent.
Equivalence would follow provided we include some Wess-Zumino-Witten
terms
\ref\witten{E. Witten, Comm. Math. Phys. 92 (1984) 455} generated by
the
quantum measure.
If we want the local Lagrangians \goryyy\ and \duall\ to be
equivalent, we must
find the conditions on $g$, $b$, $V$, $\mu$ in order to cancel the
anomalies.
To better understand the origin of the anomalies, consider first the
simpler
case where we ignore the manifest (1,0) supersymmetry and take
$b_{ij} = 0$
as well. Hence ${\cal A}_{-} = A_{-}$, and ${\cal A} = i \theta
A_{+}$.
We will show later how the formulae are modified to go back to the
general
case.

Under these simplifications the kinetic term for the fermions is:
\eqn\kinetic{ig_{ij}\lambda^i{\cal D}_{-} \lambda^j ,}
where
$${\cal D}_{-} \lambda^j = (\partial_{-} \delta^{i}\,_{j} +
\Gamma^{i}\,_{jk}
\partial_{-}x^k - A_{-} \Omega^{i}\,_{j})\lambda^j , $$
$$\Omega_{ij} = {1\over 2}(\nabla_i k_j - \nabla_j k_i). $$
Note that
\eqn\comm{[{\cal D}_{+},{\cal D}_{-}]^i \,_j =
R^i\,_{jkl}\partial_{+}x^k
\partial_{-}x^l - F_{+-} \Omega^i\,_j ,}
where
$$F_{+-} = \partial_{+}A_{-} - \partial_{-}A_{+} .$$
Working in orthonormal frames, $\delta_{ab}e^a\,_i e^b\,_j = g_{ij}$
,
\kinetic\
becomes:
\eqn\ff{i \delta_{ab} \lambda^a {\cal D}_{-} \lambda^b = i \lambda_a
( \partial_{-} \delta^a\,_b + \omega_{-}\,^a\,_b - A_{-}
\Omega^a\,_b)\lambda^b
,}
where
$\omega_{-}\,^a\,_b = \omega_i\,^a\,_b \partial_{-} x^i $ is the
pull-back of the target space spin connection to the world-sheet
and $\lambda^a = e^a\,_i \lambda^i$.

The variation of $\lambda^a $ under the gauged isometry is
\eqn\varis{\delta_{\epsilon} \lambda = - (\kappa_{L} \lambda)^a ,}
where
$$\kappa_{L}\,^a\,_b = \epsilon (k^i \omega_i + \Omega)^a\,_b .$$
Defining the effective $SO(D)$ gauge field
\eqn\effg{V_{-}\,^a\,_b = \omega_{-}\,^a\,_b - A_{-} \Omega^a\,_b ,}
\ff\ is invariant under
\eqn\dlan{\eqalign{&\delta \lambda = -\kappa_{L} \lambda ,\cr
&\delta V_{-} = \partial_{-} \kappa_{L} +[V_{-},\kappa_{L}] .\cr}}
However, the fermionic effective action
\eqn\fereff{i \Gamma^{L}\,_{eff} [V_{-}] = \log {\det}^{1\over 2}
{\cal
D}_{-}(V)}
is anomalous under \varis. The determinant \fereff\ can be computed
along the
lines
of \ref\ppoliwi{A.M. Polyakov and P. Wiegmann, Phys. Lett. B131
(1983) 121}
 in terms of the
Wess-Zumino-Witten Lagrangian for the field $g$ defined by $ V_{-} =
g^{-1}\partial_{-} g $.
Then the variation of \fereff\ under \varis\ is
\eqn\variss{\delta \Gamma^{L}\,_{eff}[V_{-}] = - {1\over 4 \pi} \int
Tr V_{-}
\partial_{+} \kappa_{L} .}

A similar computation can be carried out for the $\psi$ fermions.
The corresponding quadratic term in the Lagrangian is:
\eqn\qq{L = i \psi^{T} (\partial_{+} + V_i \partial_{+} x^i - A_{+}
\mu) \psi
.}
Defining
\eqn\ddd{V_{+} = V_i \partial_{+} x^i - A_{+} \mu ,}
we obtain an effective action:
\eqn\ee{i \Gamma^{R}_{eff}[V_{+}] = - {1\over 4 \pi} \int Tr
\partial_{-}
\kappa_{R} V_{+} .}
Adding up the two effective actions, \ee\ and \fereff, we arrive at
the
following
result,
\eqn\tt{\eqalign{&\delta \Gamma_{eff} = -{1\over 4\pi} \int
Tr(\omega_i
\partial_{-} x^i
- A_{-} \Omega)\partial_{+}(\epsilon(K^j\omega_j + \Omega))d^2
\sigma\cr
&- {1\over 4\pi} \int Tr (V_i \partial_{+} x^i - A_{+}
\mu)\partial_{-}
(\epsilon(k^j V_j + \kappa)) d^2 \sigma .\cr}}

The generalization to the (1,0) supersymmetric case and to a $b_{ij}
\neq 0$ is
quite simple. First, when $b_{ij}\neq 0$ the spin connection contains
a
contribution from the torsion. Furthermore, the $\pm v_i$
contribution to
the ${\cal J_{\pm}}$ currents complete the covariant derivatives in
$\Omega_{ij} = {1\over 2}(\nabla_i k_j - \nabla_j k_i)$ to be
covariant with respect to the full connection with torsion:
$\Omega_{ij}
\rightarrow {1\over 2} (D_i k_j - D_j k_i)$. The supersymmetric
extension
follows if we use the (1,0)-WZW Lagrangians \ref\susywz{A. Giveon, E.
Rabinovici
and A. Tseytlin, Nucl. Phys.
B409 (1993) 339; K. Sfetsos and A. Tseytlin, Nucl. Phys. B415 (1994)
116;
M. Rocek, K. Schoutens and A. Sevrin, Phys. Lett. B265
(1991) 303}.
If $d^3 Z = d^2 \sigma d\theta$,
\tt\  becomes:
\eqn\ttt{\eqalign{&\delta \Gamma_{eff} = -{1\over 4\pi} \int Tr
(\omega_i
\partial_{-}
\Phi^i - {\cal A}_{-} \Omega)D(\epsilon (k^j \omega_j + \Omega)) d^3
Z\cr
&-{1\over 4\pi} \int Tr(V_i D \Phi^i -{\cal A} \Omega) \partial_{-}
(\epsilon(k^j V_j + \mu)) d^3 Z .\cr}}
Note that unless we cancel the anomaly, the dual theory will contain
non-local contributions. The anomaly \ttt\ is a mixture between
$U(1)$
and $\sigma$-model
anomalies \ref\sagn{G. Moore, P. Nelson, Phys. Rev. Lett. 53 (1984)
1519}.
The simplest way to cancel the anomaly \ttt\ is to assume that the
spin and
gauge connection match in the original theory, a condition that also
makes
space-time anomalies cancel \killing, \anomalias. In this case if
$\mu =
\Omega$
with matching quadratic Casimirs in \ttt, the anomalous variation
\ttt\ can be cancelled by a local counterterm. Depending on the case
considered
there may be other ways to cancel the anomalies which do not require
matching $\omega$ with $V$, however we have not pursued an exhaustive
analysis.

A test of the validity of the duality transformation \dualll\ is that
if
we start with a theory with matching spin and gauge connection
$\omega = V$,
\dualll\ guarantees that in the dual theory also ${\tilde \omega} =
{\tilde
V}$.
This is also important for the consistency of the model with respect
to
global world-sheet and target-space anomalies \anomalias
, and it implies that if the original theory is conformally invariant
to
$O(\alpha^{\prime})$, so is the dual theory.

If for simplicity we take $b_{ij}=0$, the original metric is
\eqn\ormet{ds^2 = k^2 (dx^{\alpha} + A_{\alpha} dx^{\alpha})^2 +
g^{tr}\,_{\alpha
\beta}dx^{\alpha}dx^{\beta},}
where
$$g^{tr}\,_{\alpha\beta} = g_{\alpha\beta} -
{k_{\alpha}k_{\beta}\over k^2},
{}~~~~
A_{\alpha} = - {k_{\alpha}\over k^2}$$
and we can choose frames:
\eqn\frames{e^0 = k(dx^0 + A_{\alpha}dx^{\alpha}), ~~~~
e^{tr}\,^{a} = e^{tr}\,^{a}\,_{\alpha} dx^{\alpha}.}
The spin connection has components:
\eqn\spin{\eqalign{& \omega^0\,_a = \partial_a \log k e^0 -
{k\over 2} F_{ab} e^{tr}\,^{b} ,\cr
& \omega_{ab} = \omega^{tr}\,_{ab} + {k\over 2} F_{ab} e^0 ,\cr}}
where $a, b = 1,..., D-1$.
For the dual theory we have
\eqn\dualmett{d{\tilde s}^2 = {1\over k^2} (d{\tilde x}^0)^2 +
g^{tr}\,_{\alpha\beta} dx^{\alpha} dx^{\beta},}
where
$${\tilde e}^0 = {1\over k} d{\tilde x}^0 , ~~~~
{\tilde e}^a = e^{tr}\,^a .$$
Including the contribution from the torsion, the total spin
connection is:
\eqn\totspin{\eqalign{& {\tilde \omega}_{oa} =
-{\partial_a k \over k^2} dx^0 -{k\over 2}
 F_{ab}e^b\,_{\mu}dx^{\mu} ,\cr
& {\tilde \omega}_{ab} = {1\over 2} F_{ab} dx^0 +
\omega^{tr}\,_{\mu ab}dx^{\mu} .\cr }}
Using the explicit forms of \dualll\ with $ A = (0,a)$, $B = (0,b)$
(as
frame indices) it is not difficult to verify that ${\tilde
\omega}_{oa} =
{\tilde V}_{oa}$ and ${\tilde \omega}_{ab} = {\tilde V}_{ab}$. Hence
the
duality transformation preserves this condition.

There is yet one more possible source of anomalies under duality if
the
original
model is (2,0) or (2,2)-superconformal invariant. In the (2,2) case
for
instance we have a $U(1)_{L} \times U(1)_{R}$ current algebra.
The manifold has a covariantly constant complex structure,
$\nabla_k J^i\,_j = 0$, \quad $ J^i\,_k J^k\,_j = - \delta^i\,_j$.
The
R-symmetry
is generated by the rotation $\delta \lambda^i = \epsilon J^i\,_j
\lambda^j$
with current ${\cal J}_{+} = i J_{ij} \lambda^i \lambda^j$.
This current has an anomaly
\eqn\doscero{\partial_{-} {\cal J}_{+} = -{1\over 4\pi} R_{ijk}\,^l
J_l\,^k
\partial_{+}x^i \partial_{-} x^j ,}
which can be removed only if the right-hand side of \doscero\ is
cohomologically trivial. From \rv\ we know that T-duality preserves
$N=2$
global
supersymmetry (cf. also \ref\hassan{S.F. Hassan, ``O(d,d;R)
Deformations
of Complex Structures and Extended Worldsheet Supersymmetries", Tata
preprint
TIFR-TH-94-26, hep-th/9408060, August 1994; ``T-duality and Non-local
Supersymmetries", preprint CERN-TH/95-98, hep-th/9504148, April
1995}),
hence, we should be able to improve the dual R-current so that the
$U(1)_{L}
\times U(1)_{R}$ current algebra is preserved, as needed for the
application of
the
theorem in \banks. Since generically a T-duality transformation
generates a
non-constant dilaton, the energy-momentum tensor of the dual theory
contains
an improvement term due to the dilaton $S$ of the form
$\partial_{+}^2 S$.
As a consequence of $N=2$ global supersymmetry there should also be
an
improvement term in the fermionic currents and in the $U(1)$
currents. Since
the one-loop $\beta$-function implies (in complex coordinates)
$R_{\alpha{\bar \beta}} \sim \partial_{\alpha}\partial_{{\bar \beta}}
S$
\ref\lots{C.M. Hull, A. Karlhede, U. Lindstrom and M. Rocek, Nucl.
Phys. B266
(1986) 1;
T. Curtright and C. Zachos, Phys. Rev. Lett.
53 (1984) 1799}, we can improve
the $U(1)_{L} \times U(1)_{R}$ currents so that they are chirally
conserved. In the (2,2) case the improvements are:
\eqn\imp{\eqalign{& \Delta{\cal J}_{+} = \partial_{\alpha}S
\partial_{+}
Z^{\alpha} - \partial_{{\bar \alpha}} S \partial_{+} Z^{{\bar
\alpha}}, \cr
& \Delta{\cal J}_{-} = -(\partial_{\alpha} S \partial_{-} Z^{\alpha}
- \partial_{{\bar \alpha}} S \partial_{-} Z^{{\bar \alpha}}). \cr}}
For instance in the example \metrica ,   \dualmet\
the complex coordinates for the metric \dualmet\ are
\eqn\comco{z = {1\over 2} \rho^2 + i {\tilde \theta}}
and
$$ ds^2 = {dz d{\bar z}\over z +{\bar z}} .$$
Then:
\eqn\uf{\Delta{\cal J}_{+} = - {i\over \rho^2} \partial_{+} {\tilde
\theta},
{}~~~~
\Delta {\cal J}_{-} = {i\over \rho^2} \partial_{-} {\tilde \theta}.}

With these improvements the currents are chirally conserved to order
$\alpha^{\prime}$ (and presumably to all orders, since the higher
loop
counterterms are cohomologically trivial for a (2,2) supersymmetric
$\sigma$-model), hence we conclude that under duality the (2,0) or
(2,2)
superconformal algebra is preserved \foot{Hence we meet the
conditions
to apply the theorem, in \banks\ implying that the theory is
space-time
supersymmetric. More on this in section 4.}. As we will argue later
in more
detail,
the direct correspondence of operators under T-duality may map local
into
non-local operators, and
the structure of the associated low energy effective action has to be
understood with some care. The mapping of local into non-local states
is familiar from the case of toroidal duality, where momentum states
are
mapped into winding states.

\newsec{The effective action point of view}
In this section we want to give an answer to some of the puzzles
raised in BKO
\tsusy\ and analyze the example \metrica, \dualmet\  from the
effective
action point of view.
The original background describes the motion of the heterotic string
in flat
Minkowski space. Hence we have full $ISO(1,9)$ Lorentz invariance and
$O(32)$
gauge symmetry (the same arguments apply to the $E_8 \times E_8$
string).
Since we perform duality in \metrica\ with respect to rotations in
the $(x,y)$
plane, only the subgroup of $ISO(1,9)$ commuting with them will be a
manifest
local symmetry of the effective action. Similarly if we preserve
manifest (1,0)
supersymmetry on the world-sheet and avoid anomalies, we embed the
isometry
group $SO(2) \subset G \equiv SO(32)  $. The subgroup of $G$
commuting with
$SO(2)$ is
$SO(30) \times SO(2)$ and once again this will be a manifest symmetry
in the low energy theory. It is well known \kiritsis, \tdual,
\canonical\ that
under T-duality, symmetries not commuting with the ones generating
duality
 are generally realized non-locally. Hence although the dual
background
\dualmet\ still contains all the original symmetries from the CFT
point of
view,
the low energy theory does not seem to exhibit them. The theory will
be
explicitly symmetric under $ISO(1,7) \times SO(30) \times SO(2)$
only.
We want to make sure nevertheless that the original space-time
supersymmetry
is preserved.

The world-sheet supersymmetry \superc\ commutes with (1,0)
T-duality, and, from the arguments in the previous section we expect
the
dual theory to exhibit the equivalent of the full $N=1$ space-time
supersymmetry of the original space, although not necessarily in a
manifest $O(1,9)$-covariant formalism. To find the graviton,
gravitino, etc
vertex operators in the dual background \dualmet\ we should solve the
anomalous
dimension operators constructed in the metric \dualmet\ including the
dilaton and background gauge field, as was done for tachyons in
\ref\cg{C. Callan and Z. Gan, Nucl. Phys. B272 (1986) 647}. We can
proceed
differently.
The dual background is
\eqn\dualb{\eqalign{& d{\tilde s}^2 = d\rho^2 +{1\over \rho^2}
d{\tilde \phi}^2
-
(dx^0)^2 +(dx^i)^2 \ ; ~~~ i = 1,...,7 , \cr
& S = - \log \rho , ~~~~~
V_{\mu} dx^{\mu} = {1\over \rho^2} d{\tilde \phi} M , \cr}}
where $M$ is the matrix describing the embedding of the spin
connection
in the gauge group, which we take to be the standard one acting only
on two
of the right-moving fermions. \dualb\ satisfies the heterotic
$\beta$-function
equations \betaa\ to $O(\alpha^{\prime})$ . We can consider the
variation
of the fermionic degrees of freedom in a formalism adapted to the
$ISO(1,7)
\times SO(30) \times SO(2)$ symmetry, and look for which combination
of
the $O(1,9)$ fermions are annihilated by supersymmetry. We believe
these
combinations
are the ones that we would obtain if we followed the  method
in \cg . We will find that the number of space-time fermionic
symmetries
does not change.
\par
The low energy approximation to the heterotic string is given by
$N=1$ supergravity coupled to $N=1$ super Yang-Mills in $d=10$. In
ten
dimensions
we can impose simultaneously the Majorana and Weyl conditions
\ref\spinor{L.
Alvarez-Gaume, ``An Introduction to Anomalies''
(in: ``Erice School in Mathematical Physics'', Erice, 1985); P. van
Nieuwenhuizen,
``An Introduction to Supersymmetry, Supergravity and the Kaluza-Klein
Program''
(Les Houches, 1983; North Holland, 1984)}. In terms of $SO(1,7)$,
a Majorana-Weyl spinor of $SO(1,9)$ becomes a Weyl spinor. Write the
Dirac
algebra (in an orthonormal frame) as
\eqn\dirac{\eqalign{& \Gamma_{\mu} = \tau_3 \otimes \gamma_{\mu}\ ;
{}~~~
\mu = 0,1,...,7 , \cr
& \Gamma_{7+i} = i \tau_i \otimes 1\ ; ~~~ i =1, 2 , \cr
& {\bar \Gamma} = \tau_3 \otimes \gamma_9 , \cr}}
where $\gamma_9$ is the analogous of the four-dimensional $\gamma_5$
in eight dimensions. Ten dimensional
indices will be hatted.
The supersymmetric variation of the
ten-dimensional fermions is given by:
\eqn\fervar{\eqalign{& \delta{\hat \Psi}_{{\hat \mu}} =
(\partial_{{\hat \mu}} - {1 \over 4}
\omega_{{\hat \mu}{\hat a}{\hat b}} \Gamma^{{\hat a}{\hat b}}){\hat
\epsilon},
\cr
& \delta {\hat \lambda} = (\Gamma^{{\hat \mu}} \partial_{{\hat \mu}}
S -
{1\over 6} H_{{\hat \mu}{\hat \nu}{\hat \rho}} \Gamma^{{\hat
\mu}{\hat \nu}
{\hat \rho}}){\hat \epsilon}, \cr
& \delta \chi^A = -{1\over 4} F^A\,_{{\hat \mu}{\hat \nu}}
\Gamma^{{\hat \mu}{\hat \nu}}{\hat \epsilon} \cr}}
for the gravitino, dilatino and gluino, respectively. The background
gauge
field
strength is:
\eqn\bgs{F = -{2\over \rho^3}d\rho \wedge d{\tilde \phi}.}

Decomposing \fervar\ with respect to $SO(1,7)$ we find:
\eqn\dfe{\eqalign{& \delta \Psi_{\mu} = \partial_{\mu} \epsilon, \cr
& \delta \Psi_{\{\rho\}} = \partial_{\{\rho\}}\epsilon , \cr
& \delta \Psi_{\{{\tilde \phi}\}} = (\partial_{\phi} + {i \over
4\rho^2}
\tau_3 \otimes 1)\epsilon \cr}}
for the gravitino
\foot{We have represented by $\Psi_{\{\rho\}}$ and
$\Psi_{\{\tilde{\phi}\}}$
the components of $\Psi_{\mu}$
when $\mu = \rho$ and $\mu = {\tilde \phi}$, respectively.},
\eqn\grti{\delta \lambda = -{ i\over \rho} (\tau_1 \otimes 1)
\epsilon}
for the dilatino, and
\eqn\aa{\eqalign{& \delta \chi^A = 0 ; ~~~~ A \in SO(30), \cr
& \delta \chi =- {i\over \rho^2} (\tau_3 \otimes 1)
\epsilon ;~~~~along~~the~~ embedded~~ SO(2),\cr}}
for the gluino. In the preceding formulas, $\epsilon$ is now an
$SO(1,7)$ Weyl
spinor with the same number of independent components as a
ten-dimensional
Majorana-Weyl spinor.

Next, if we define
\eqn\des{\eqalign{&{\tilde \Psi}_{\mu} = \Psi_{\mu} ,\cr
&{\tilde \Psi}_{\{\rho\}} = \Psi_{\{\rho\}} ,\cr
&{\tilde \Psi}_{\{{\tilde \phi}\}} = \Psi_{\{{\tilde \phi}\}}
+ {i \over 4}e^{S}  (\tau_2 \otimes 1) \lambda ,\cr}}
it is easy to see that the fields on the left hand side transform as
$\delta {\tilde
\Psi}_{{\hat \mu}} = \partial_{{\hat \mu}} \epsilon$. Similarly,
\eqn\dess{\eqalign{&{\tilde \lambda} = \lambda  + i e^{-S} (\tau_2
\otimes 1)
\chi ,\cr
&{\tilde \chi} = \chi - i e^S (\tau_2 \otimes 1) \lambda\cr}}
have vanishing variation under space-time supersymmetry. Furthermore,
they
have the correct chiralities as dictated by the ten-dimensional
multiplet.
\par
Thus if we use a formalism covariant only under the explicit $SO(1,7)
\times
SO(30) \times SO(2)$ symmetry of the background
\dualb\ we recover the full number of supersymmetric charges. From
the
low-energy
effective action this is the most we could expect since at the level
of
the world-sheet CFT the full symmetry $SO(1,9) \times SO(32)$ is only
realized
non-locally. If we want to consider the complete symmetry and the
complete
massless spectrum in the dual theory it seems that the only
reasonable thing
to do is to go back to the two-dimensional point of view. This is not
unreasonable because at $\rho =0$ the dual background has a curvature
singularity and the naive $\sigma$-model and effective action
considerations
should not be trusted. We will see in the next section how one can
obtain in principle the vertex operators for the full massless
spectrum in
the dual theory.
\par
An interesting exercise would be to verify that indeed \des , \dess\
are the
vertex operators one would obtain for the dual background \dualb\ if
we
followed reference \cg .
\newsec{Conformal field theory analysis}
As suggested in the introduction, a representative example of duality
transformations with respect to isometries having fixed points is to
analyse
the heterotic string in flat ten-dimensional Minkowski space and
perform
duality with respect to rotations.
This example can be studied explicitly in detail using conformal
field
theory (CFT) techniques, and there are a number of things that can be
learned.
In the previous section we analyzed this example from the low-energy
effective action point of view, and we have learned that only part of
the
 symmetries of the original theory  manifest locally in the dual
transform.
In this section we want to investigate in more detail the way the
full
symmetry is realized.

If we perform duality in the $(x,y)$-plane, the part of the free
heterotic
Lagrangian of interest is
\eqn\cuauno{L = \partial_{+}{\vec x} \cdot \partial_{-}{\vec x} +
i {\vec \lambda} \cdot \partial_{-}{\vec \lambda} +
i \psi^A \partial_{+} \psi^A +... ,}
where the vector quantities are two-dimensional.
The isometry we consider is
\eqn\cuados{x \rightarrow e^{\epsilon \alpha} x ,}
where
$$ \epsilon = i \sigma_2 = \left(\matrix{0&1\cr
-1&0\cr}\right).$$
For the time being we work in frames not adapted to the isometry.
Hence
for \cuauno\ we can perform duality only in the bosonic sector.
The world-sheet supercurrent is
\eqn\cuatres{{\cal G}_{+} = {\vec \lambda} \cdot \partial_{+} {\vec
x} = {\vec
\lambda}
\cdot {\vec P}_{+},}
where ${\vec P}_{+}$ is a chiral current generating translations in
the
target space.

It is convenient to work in canonical pictures \picture\ ($-{1\over
2}$
for fermion vertices, $-1$ for boson vertices).
The space-time supersymmetry charge is
\eqn\cuacua{Q_{\alpha}\,^{(-{1\over 2})} = \oint e^{-{\phi \over 2}}
S_{\alpha},}
where $\phi$ is the scalar which bosonizes the superconformal ghost
current
and $S_{\alpha}$ is the spin-field associated to the
$\lambda$-fermions.
The translation operator in the $-1$ picture is
\eqn\cuacin{P_{\mu}\,^{(-1)} = \oint e^{-\phi} \lambda_{\mu}.}
Note that in \cuatres,  \cuacua\ only the space-time fermion and the
$(\beta ,\gamma)$-ghosts appear. Hence
\eqn\cuasei{\{Q_{\alpha}\,^{(-{1\over 2})} , Q_{\beta}\,^{(-{1\over
2})} \}
= \Gamma^{\mu}P_{\mu}\,^{(-1)}}
is satisfied, and if we choose to perform duality for the bosonic
part of
the Lagrangian
only, the same relationships \cuacua, \cuacin, \cuasei\ should still
hold.

{}From this point of view there is clearly no problem with space-time
supersymmetry.
However, in constructing scattering amplitudes we need to use
vertex operators in different pictures. Hence any problem should
come from the interplay with the picture changing operator ${\cal
P}$.
The picture changing operator acting on a vertex operator $V_q (z)$
in the
$q$-picture  can be represented as \picture
\eqn\cuasiete{{\cal P} V_q (z) = lim_{w \rightarrow z} e^{\phi(w)}
{\cal G}_{+}(w) V_q(z) .}
Hence the only possible difficulties may appear in anomalies in the
world-sheet
supercurrent under duality. Since ${\cal G}_{+}$ does not commute
with
purely bosonic rotations, after duality ${\cal G}_{+}$ will become
non-local
in the world-
sheet.
To guarantee that there are no problems with ${\cal G}_{+}$ we want
to make
sure
that the dual world-sheet supercurrent still has the form ${\vec
\lambda} \cdot
{\tilde {\vec P}}_{+}$, where ${\tilde P}_{+}$ is the representation
of the
translation current in the dual theory, and it is here that
non-locality resides. In fact the full theory in \cuauno\ can be
constructed
out of the knowledge that $P_{+}\,^i$ ($i = 1, 2$) is chirally
conserved
and that its operator product expansion (OPE) is
$P^i (z) P^j (w) \sim {\delta^{ij} \over (z - w)^2}$. It is hard to
believe that the existence of the chiral currents is going to be lost
under duality. To make sure that this is not the case, the simplest
thing to do is to include sources for these currents and then follow
their transformation under duality.

Following \rv \foot{There are two procedures presented in \rv . One,
which
we follow, is to gauge the isometry and impose the vanishing field
strength
constraint. The second procedure consists of writing a $\sigma$-model
in $D+1$ dimensions, where $D$ is the dimension of the original
manifold
so that the original isometry is promoted to a full $U(1)_L \times
U(1)_R$
Kac-Moody algebra. Then the original and the dual theories are
obtained
by gauging respectively vector or axial vector combinations of the
two $U(1)$.
This procedure however has to be modified to include also a dilaton
in $D+1$
 dimensions to preserve conformal invariance (as can be easily
checked
by working out the $\beta$-function equations), and to guarantee that
the
dilaton
transforms as in \buscher, \ref\ao{E. Alvarez and M.A.R. Osorio,
Phys. Rev.
D40 (1989) 1150} under duality. Since we do not want
to keep track of extra dilaton contributions, we follow the first
procedure.}
we gauge the symmetry \cuados\ and concentrate only on the bosonic
part of
\cuauno, the only one relevant due to the previous arguments.
Thus our starting point is:
\eqn\cuaocho{L = D_{+} x^{T}  D_{-} x + \lambda F_{+-},}
where
$$D_{\pm} x = \partial_{\pm} x + \epsilon x A_{\pm}, ~~~~
F_{+-} = \partial_{+} A_{-} - \partial_{-}A_{+} . $$
Using the $\lambda$-equation of motion, $A_{\pm} =
\partial_{\pm}\alpha$,
$D_{\pm} x = e^{-\epsilon\alpha}
\partial_{\pm}(e^{\epsilon\alpha}x)$, and
changing variables $ x \rightarrow e^{-\epsilon \alpha} x$, the
original theory is recovered.
It proves convenient to parametrize locally
\eqn\cuanueve{A_{+} = \partial_{+} \alpha_L , ~~~~~
A_{-} = \partial_{-} \alpha_R . }
Then \cuaocho\ has the symmetries:
\eqn\cuadiez{\eqalign{&\delta x = e^{-\epsilon \alpha_{R}} a_R ,\cr
&\delta \lambda
= - x^T \epsilon e^{-\epsilon\alpha_R} a_R , \cr}}
and
\eqn\otherp{\eqalign{&\delta x = e^{-\epsilon\alpha_L} a_L , \cr
&\delta \lambda = x^T
\epsilon e^{-\epsilon \alpha_L} a_L ,\cr}}
yielding respectively the conserved currents:
\eqn\cuaonce{\partial_{-}(e^{\epsilon\alpha_R} D_{+} x) = 0 , ~~~~
\partial_{+}(e^{\epsilon \alpha_L} D_{-} x) =0 .}
When $\alpha_R = \alpha_L = \alpha$, we recover the original currents
$\partial_{\pm}(e^{\epsilon\alpha} x)$. Acting on $x$, the symmetries
\cuadiez\
commute.

The sources to be added to \cuaocho\ should be gauge invariant:
\eqn\cuadoce{J_{-} e^{\epsilon\phi_R} D_{+} x + J_{+} e^{\epsilon
\phi_L} D_{-}
x .}
The exponents are non-local in $A_{\pm}$, and they make the coupling
gauge invariant. This also guarantees that the currents in \cuadoce\
satisfy
the OPE of the original theory as expected. The coupling \cuadoce\ is
quite different from the classical toroidal model where duality
is performed with respect
to translational symmetries. In this case the analogue of \cuaocho,
\cuadoce\
is
\eqn\cuatrece{R J_{-} D_{+} x + R J_{+} D_{-} x + R^2 \partial_{+} x
\partial_{-} x + \lambda F_{+-},}
where
$$D_{\pm} x = \partial_{\pm} x + A_{\pm} .$$
After duality, we obtain:
\eqn\cuacator{{1\over R^2} \partial_{+} \lambda \partial_{-}\lambda
+ {1\over R} (\partial_{+}\lambda J_{-} - \partial_{-}\lambda J_{+})
+
\partial_{-} J_{+}
{1\over \partial_{+}\partial_{-}} \partial_{+} J_{-},}
exhibiting quite clearly the exchange between momentum and winding
modes.
In our example, let us take for simplicity $J_{+} = 0$. The most
straightforward
way to integrate out the gauge fields is to work in the light-cone
gauge
 $A_{-} = 0$. Then the integral over $A_{+}$ becomes a
$\delta$-function
which can be solved in two ways. If we choose to solve it in order to
write
the Lagrange multiplier $\lambda$ as a function of the other fields,
we
recover the original theory.

On the other hand if we choose to solve the adapted coordinate to the
isometry
in terms of $\lambda$ and the other variables, we obtain the dual
theory.
Furthermore we also obtain a determinant, which when properly
evaluated
\buscher, \tdual\ yields the transformation of the dilaton.
We use complex coordinates,
\eqn\cuaquin{z = x+ i y = \rho e^{i \theta} ,}
and
\eqn\alex{J_{-} = J_{-}\,_1 + i J_{-}\,_2 , }
where $J_{-}\,_{1, 2}$ are the components of $J_{-}$ in \cuadoce ;
the
complex conjugate of $J_{-}$ will be denoted by $J_{-}^{*}$. Then,
we obtain the equation
\eqn\cuadseis{\partial_{-} w = w {i\over \rho^2} \partial_{-} \lambda
- {\rho\over 2 i}(J_{-} - J_{-}^{*} w^2),}
where
$$w = e^{i \theta} ,$$
ie. a Riccatti equation which can be solved order by order in
$J_{-}$.
Restoring powers of $\alpha^{\prime}$,
\eqn\cuadsiete{\partial_{-}w = i w {\alpha^{\prime}\over
\rho^2}\partial_{-}
\lambda - {\rho \over 2 i \alpha^{\prime}}(J_{-} - J_{-}^{*} w^2).}

The lowest order solution is:
\eqn\cuadocho{ w = \exp i \int {\alpha^{\prime} \over
\rho^2}\partial_{-}
 \lambda d\sigma^{-}}
and to this order the current looks like
\eqn\cuadnueve{\partial_{+} \left(\rho e^{i \theta[\lambda,J_{-}]}
\right) = \partial_{+}\left(
\rho e^{i\alpha^{\prime}\int {1\over \rho^2}\partial_{-}\lambda
d \sigma^{-}}\right).}
The extra terms depending on $J_{-}$, $J_{-}^{*}$ are required
to guarantee the equality between the correlation functions before
and
after the duality transformation.
To leading order the currents are:
\eqn\cuadd{\eqalign{&\partial_{+} \left( \rho e^{\pm i
\alpha^{\prime}
\int {1\over
\rho^2} \partial_{-} \lambda d \sigma^{-}} \right),\cr
&\partial_{-} \left(\rho e^{\pm i \alpha^{\prime} \int {1\over
\rho^2}\partial_{+}
\lambda d\sigma^{+}} \right) .\cr}}
Note however that in solving \cuadsiete\ there will be corrections to
all
orders in $\alpha^{\prime}$ in order to obtain the correct OPE's for
dual currents ${\tilde P}_{\pm}$ . These currents can be used to
write the emission vertex operators in the dual theory and they are
almost
always non-local. Since the OPE's of ${\tilde P}_{\pm}$ are
preserved, the
spectrum of the original and the dual theories are equivalent.
Nevertheless, we
have to be careful regarding the
operator mapping.

In the original theory the target space has a flat metric
while in the dual manifold \dualmet\ we have curvature singularities
at $\rho = 0$ . In the original theory the vertex operators are
expressed in
terms of momentum states, which in the dual theory correspond
to winding states. This is reasonable because operators like \cuadd\
would
produce non-normalizable states acting on momentum states, but
normalizable acting on winding states on the dual space, as one can
easily
show by looking at the solutions of the anomalous dimension operators
on the
dual
background \cg.
We also see that when there are curvature singularities the
$\sigma$-model
approach does not work. We can only trust this approximation when
the manifold is large, or the curvature is small when expressed
in units of $\alpha^{\prime}$. Since in the dual theory near
$\rho = 0$ we have a curvature singularity, the low energy field
theory limit cannot be trusted close to this point, and we have to
deal with
the dual theory without relying on an expansion in powers of
$\alpha^{\prime}$.
If there were  a minimum value $\rho_{min} \neq 0$ we would still
be able to arrange the solution of \cuadsiete\ in a power series of
$\alpha^{\prime}$.
When this is not the case, we can only obtain reliable information
if we can control the exact operator mapping. Equations \cuadsiete
-\cuadd\
show that in order to recover the expected OPE for the chiral
currents
${\tilde P}_{\pm}$ we need to include all orders in
$\alpha^{\prime}$, which
then in
turn imply that the world-sheet supercurrent and the picture changing
operator
have the expected properties.
\par
In conclusion, there is no problem with space-time supersymmetry from
the
 point of view of CFT, but the correct operators that need to be used
to represent the emission vertices of low energy particles in the
dual
theory are often non-local, and do not admit a straightforward
$\alpha^{\prime}$
expansion unless we write the dual states in terms of those which
follow from the correspondence as dictated by the duality
transformation.
When there are curvature singularities the approach based on the
effective
low energy theory has many limitations and to obtain reliable
information we
should go back to the underlying string theory.

\newsec{Marginal deformations and supersymmetry}

The issue of space-time supersymmetry versus duality arises even
infinitesimally, when considering $O(d,d)$ deformations of
superconformal
backgrounds with $d$ abelian isometries. If some of the Killing
vector
fields have fixed points in their action, the corresponding
deformations
will exhibit an ``anomalous" supersymmetric behaviour in a way
similar to
the previous analysis. It is particularly interesting in this context
to
examine the special class of $O(d,d)$ transformations that describe
marginal $J \bar{J}$-deformations of a given conformal field theory,
although the generalization to all other $O(d,d)$ elements can be
easily
incorporated. The most characteristic class of examples for our
present
purposes is provided by the $N=2$ superconformal string backgrounds
on
group manifolds $G$, or symmetric coset spaces, which are
known as Kazama-Suzuki models \Kazama . We will
investigate the effect of a $J \bar{J}$-deformation on the
supersymmetric
properties of the associated $G_{k}$ WZW models, with main emphasis
on
$G \equiv SU(2) \times U(1)$ as the simplest but representative
example that
contains most of the difficulties encountered in the general case.

We have typically the following deformation of the $\sigma$-model
action

\eqn\defo{S_{R+ \delta R} = S_{R} + {{\delta R}^{2} \over 4 \pi k}
\int J(R) \bar{J}(R) ,}
where $R$ is a modulus field and $S_{R=1}$ is the corresponding
action of the
undeformed WZW model. The main point that makes relevant our previous
analysis to the present case is the realization that all
Kazama-Suzuki group
manifolds admit some Killing vector fields with fixed points, which
in turn
enter into the description of the $J \bar{J}$-deformations as
$O(d,d)$
transformations. The undeformed models at $R=1$ admit a Killing
spinor and
hence they have manifest $N=1$ space-time supersymmetry. The Killing
spinors,
however, depend explicitly on the (adapted) coordinates of the
corresponding
rotational Killing vector fields, in exact analogy with the
$\theta$-dependence
of the Killing spinors in flat space with polar coordinates $(\rho,
\theta)$.
Switching on the $J \bar{J}$-deformation, we find that there are no
solutions
to the Killing spinor equation neither infinitesimally away from
$R=1$ nor
at the end points (say $R=0$ and $\infty$), the latter being related
to each
other by factorized duality. Also, away from $R=1$, the dilatino
variation
is not zero, as it would have been required from the supersymmetric
properties
of a bosonic solution, and it appears as if space-time supersymmmetry
is lost
without including any appropriate world-sheet effects.

There is an interesting
phenomenon that we will find at this end for the simplest background
of this
type, $SU(2) \times U(1)$. Namely, there is a dynamical restoration
of
space-time supersymmetry as soon as we are prepared to alter our
model by
making the modulus field $R$ into a dynamical variable, while
preserving
conformal invariance. We will present some details of this mechanism
,
since we believe that it is of general value and deserves systematic
study
for all Kazama-Suzuki models and perhaps further beyond. This
phenomenon
seems to indicate that there is a systematic way to include
dynamically all the
appropriate non-perturbative world-sheet effects into a new string
vacuum
with manifest space-time supersymmetry.

We consider for simplicity the $SU(2)$ WZW model in the following and
parametrize
its group elements as

\eqn\groupel{g = e^{i(\tau - \psi) \sigma_{3} / 2} e^{i \varphi
\sigma_{2}}
e^{i(\tau + \psi) \sigma_{3} / 2} .}
The action in these coordinates describes a 2-dim $\sigma$-model with
target
space
metric

\eqn\linele{{ds}^{2} = {\sin}^{2} \varphi {d \psi}^{2} + {\cos}^{2}
\varphi
{d \tau}^{2} + {d \varphi}^{2}}
and anti-symmetric tensor field with non-zero component

\eqn\anti{b_{\tau \psi} = {\cos}^{2} \varphi ,}
whereas the dilaton field $S$ is zero.

The $J \bar{J}$-deformation of the $ SU(2) \times U(1)$ WZW model can
be
easily induced from the corresponding deformation of the $SU(2)$
model. The
$R$-line of deformed $SU(2)$ models is known to have the following
structure \ref\Amit{S.F. Hassan and A. Sen, Nucl. Phys. B405 (1993)
143;
A. Giveon and E. Kiritsis, Nucl. Phys. B411 (1994) 487}:

\eqn\linee{{ds}_{(R)}^{2} = {1 \over {\cos}^{2} \varphi + R^{2}
{\sin}^{2} \varphi} \left({\sin}^{2} \varphi {d \psi}^{2} + R^{2}
{\cos}^{2} \varphi {d \tau}^{2} \right) + {d \varphi}^{2} ,}

\eqn\antii{b_{\tau \psi}^{(R)} = {{\cos}^{2} \varphi \over {\cos}^{2}
\varphi
+ R^{2} {\sin}^{2} \varphi}}
and

\eqn\dila{S^{(R)} = {1 \over 2} \log {R \over {\cos}^{2} \varphi
+ R^{2} {\sin}^{2} \varphi} .}
We note for completeness that the conformal invariance along the
$R$-line of
marginal deformations requires that
$e^{-2 S^{(R)}} \sqrt{\det g(R)}$ remains the same for all
$0 \leq R < \infty$ in the $\sigma$-model frame. Also, for $R=1$, we
readily
obtain the original $SU(2)$ WZW model.

The $J \bar{J}$-deformation of the $SU(2) \times U(1)$ WZW model can
be simply
performed by adding an abelian $U(1)$ field $\rho$ with the
appropriate
background
charge, ie. a linear dilaton. Then, the resulting 4-dim backgrounds
along the
$R$-line have the following metric, anti-symmetric tensor and dilaton
fields
respectively,
\eqn\datum{{d \rho}^{2} + {ds}_{(R)}^{2}, ~~~~ b_{\tau \psi}^{(R)},
{}~~~~
- \rho + S^{(R)} .}
We note at this point that although at $R=1$ the $SU(2) \times U(1)$
WZW model
has manifest space-time supersymmetry
\foot{This model actually has $N=4$
world-sheet supersymmetry and $N=2$ space-time supersymmetry.},
the supersymmetry appears to be broken even infinitesimally away from
it. The
simplest way to see this is by considering the dilatino variation at
generic
values
of $R$. The vanishing condition $\delta \lambda = 0$ would imply the
balance
\eqn\balan{{1 \over 2} H_{\mu \nu \rho} = - \sqrt{\det g} ~
{{\epsilon}_{\mu \nu \rho}}^{\sigma} {\partial}_{\sigma} S}
in the $\sigma$-model frame of the 4-dim class of models \datum\ .
Straightforward
calculation yields
\eqn\cond{{\cos}^{2} \varphi + R^{2} {\sin}^{2} \varphi = R ,}
which is clearly satisfied only for $R=1$, as advertised earlier.

The reason for this occurence can also be traced to the specific form
of the
Killing spinors at the $R=1$ point. We have determined that the
spinor with
components \foot{We thank K. Sfetsos for supplying this formula.}

\eqn\spinor{\xi_{\pm} = e^{i \varphi \sigma_{3} / 2}
e^{i (\psi \mp \tau) \sigma_{1} / 2} \epsilon_{\pm}}
solves the Killing spinor equation for the $SU(2) \times U(1)$
background,
where
$\epsilon_{\pm}$ are the Weyl components of a constant Killing
spinor.
According to our general framework, it is the existence of fixed
points in the
action of the Killing vector fields $\partial / \partial \psi$ and
$\partial / \partial \tau$ that accounts for the $\psi$ and $\tau$
dependence
of the Killing spinor \spinor\ . The $J \bar{J}$-deformation in
question is
generated by an appropriately chosen $O(2,2)$ transformation that is
associated with these two Killing vector fields and hence the loss of
space-time supersymmetry comes as no surprise in this case. As
before, the
inclusion of non-perturbative world-sheet effects could resolve the
issue of
supersymmetry versus duality, although it might not necessarily be so
at the
infinitesimal level we are considering here.

We turn now to the detailed description of a mechanism that restores
dymanically the space-time supersymmetry and yields a new 4-dim
string
background based on $SU(2)$. Recall that in our previous discussion
the
field $\rho$ was added as a fourth space-time coordinate, but it was
important that it remained a spectator, in the sense that the modulus
field
$R$ driving the deformation was $\rho$-independent. This was of
course
necessary for constructing a one-parameter family of 4-dim conformal
backgrounds with $R=1$ being the $SU(2) \times U(1)$ model. Next, we
will
abandon this parametric construction by altering the
role of the field $\rho$. We will introduce instead a single 4-dim
background having $R$ with a dynamical $\rho$-dependence, in that at
any given
instance of ``time" $\rho$ the 3-dim slices are points in the moduli
space
of $J \bar{J}$-deformations of the $SU(2)$ WZW model\foot{We choose
$\rho$ to have the signature of physical time in order to stress this
analogy. The Euclidean version can be obtained by simple analytic
continuation $\rho \rightarrow i \rho$.}. The point of this method is
mainly the observation that space-time supersymmetry can be restored
for
such a background. Although this dynamical background appears to be
disconnected from the moduli space of the $R$-deformed $SU(2) \times
U(1)$
models, it might be relevant for describing some of their
non-perturbative
aspects in $\alpha^{\prime}$. This possibility deserves further
study.

The 4-dim metric in this case is taken $-{d \rho}^{2} +
{ds}_{(R)}^{2}$,
accounting
for the $-+++$ signature, while the anti-symmetric tensor field is
given again
by \antii. The condition of conformal invariance is sufficient to
determine the
dilaton field
\eqn\dilaa{S^{(R)} = {1 \over 2} \log {R^{\prime} \over {\cos}^{2}
\varphi
+ R^{2} {\sin}^{2} \varphi} ,}
where the prime denotes the derivative with respect to $\rho$. This
model was
originally considered as paradigm for topology change in string
theory
\ref\Kounnas{E. Kiritsis and C. Kounnas, Phys. Lett. B331 (1994) 51},
where it
was
also found that $R$ satisfies the differential equation
\eqn\dynami{{R^{\prime \prime \prime} \over R^{\prime \prime}} =
{R^{\prime \prime} \over R^{\prime}} + {R^{\prime} \over R} .}
Simple integration yields
\eqn\integr{R^{\prime} = C_{1} R^{2} + C_{2} ,}
where $C_{1}$ and $C_{2}$ are arbitrary constants for the time being.

There are three classes of solutions depending on whether $C_{1} = 0$
(or
$C_{2} = 0$), $C_{1} C_{2} > 0$ or $C_{1} C_{2} < 0$. The observation
now is
that
not all of these solutions are space-time supersymmetric. Explicit
calculation
shows that only the third class exhibits space-time supersymmetry (to
lowest
order in $\alpha^{\prime}$), as it can be verified by considering the
vanishing
condition of the dilatino variation. In this case we find $C_{1} =
-C_{2} = 1$
and integration of \integr\ determines the relevant supersymmetric
solution
\eqn\final{R(\rho) = \tanh \rho}
for the dynamical model. It is known that this is T-dual to the
${SU(2) \over
U(1)}
\times {SL(2) \over U(1)}$ WZW coset model \Kounnas, which also has
manifest
space-time supersymmetry.

The method of the dynamical restoration of space-time supersymmetry
can be
generalized to any Kazama-Suzuki model, but we will not present the
details
here.
An interesting question is to determine the class of solutions that
restore supersymmetry in the general case. Also the deeper connection
between
the original $R$-family of deformed WZW models and the associated
dynamical
backgrounds remains to be clarified. We hope that this mechanism
incorporates
in
some way the appropriate world-sheet effects that have to be taken
into account
to resolve the problem of supersymmetry versus duality that arises in
the
context of the effective theory.

\newsec{Conclusions}

To summarize, we stress again that the problem that seems to arise
in the effective theory concerning the fate of space-time
supersymmetry under
T-duality with respect to rotational Killing vector fields, is
entirely due
to non-local world-sheet effects that have to be taken into account
in
string theory. Indeed, our analysis provides the resolution to this
problem
in the context of CFT, where the physically correct operators for the
emission
vertices of low  energy particles are often non-local without a
straightforward $\alpha^{\prime}$ expansion. We have rediscovered
from a
different prospective that the low energy effective theory is not
trustworthy
for the description of regions close to curvature singularities. The
mechanism
for the
dynamical restoration of space-time supersymmetry that we considered
in
the last section could be used to incorporate the world-sheet effects
that are
needed to extend the validity of the effective theory. After all, the
initial motivation for promoting
various moduli fields into dynamical variables was the
extension of the effective field
theory approach to string dynamics by patching together the regions
where
the low energy approximations are valid \Kounnas.

A very useful exercise would
be to study the anomalous dimension operators in the general (1,0)
supersymmetric $\sigma$-model along the lines of \cg.
The most straightforward and logical attack to the problem would
of course be to maintain manifest space-time supersymmetry all
along the calculation, in the Green-Schwarz formalism.
Due to our limited understanding of the quantization
of this theory, however, we are forced for the time being to use
NSR fermions.
This implies in particular that in order to relate these
$\sigma$-model calculations
to space-time supersymmetry, one has presumably to solve the
problem of properly including sources for the spin operators.

We would like to point out that in reference \ref\spencer{C. Hull,
Mod. Phys.
Lett.
A5 (1990) 1793; C. Hull and B. Spence, Nucl. Phys. B345 (1990) 493} a
similar
problem is found. They work with (1,0) and (2,0) WZW models, where
they find a
clash between the super-Kac-Moody symmetry and the second
supersymmetry. They
show that if the Kac-Moody transformations are accompanied by a
compensating
deformation of the complex structure, the problem is resolved and
yields a
(2,0) supersymmetric extension of the superconformal and Kac-Moody
algebras. This compensating transformation is not the one that
naturally
follows from T-duality in the case of rotational (non-holomorphic)
isometries. We suspect that the dynamical restoration of
supersymmetry in
section 5 is closely related to the need of having a compensating
deformation
of
the complex structure.

Some of the topics of this paper may also be relevant for
superstring phenomenology, where we mainly work in terms of the
lowest order effective theory. The issue of space-time supersymmetry
versus duality, which arises to lowest order in $\alpha^{\prime}$,
demonstrates explicitly that an apparently non-supersymmetric
background can
qualify as a vacuum solution of superstring theory, in contrary to
the ``standard wisdom" that has been considered so far. So, whether
supersymmetry is broken or not in various phenomenological
applications
cannot be decided, unless one knows how to incorpotate the
appropriate
(non-local) world-sheet effects that might lead to its restoration
at the string level. Also, various gravitational solutions, like
black-holes,
might enjoy some supersymmetric properties in the string context.
This might also provide a better understanding of the way that string
theory, through its world-sheet effects, can resolve the fundamental
problems of the quantum theory of black-holes.
We hope to return to these problems elsewhere.

Finally, from the low energy point of view the fact that T-duality
relates supersymmetric with non-supersymmetric backgrounds could
provide examples of the mechanism advocated by Witten \ref\laston{E.
Witten,
Int. J. Mod. Phys. A10 (1995) 1247}
to shed new light in the cosmological constant
problem.

\bigbreak\bigskip\bigskip\centerline{{\bf Acknowledgements}}\nobreak
The work of E.A. has been partially supported by CICYT (Spain),
 under contract AEN/93/0673.
We are grateful to Eric Bergshoeff, Elias Kiritsis,
Kostas Kounnas, Wolfgang Lerche, Tom\'as
Ort\'{\i}n,
Fernando Quevedo, Kostas Sfetsos and Erik Verlinde for many useful
discussions.
We also acknowledge the hospitality of the Benasque Center for
Physics in
Spain,
where this work was completed.

\listrefs

\bye